\documentstyle[12pt]{article}

\def\beqa{\begin{eqnarray}}
\def\eeqa{\end{eqnarray}}
\def\beq{\begin{equation}}
\def\eeq{\end{equation}}

\def\umunu{^{\mu\nu}}

\def\ddemu{_{;\mu}}  
\def\ddenu{_{;\nu}}

\def\bib#1{$^{\ref{#1}}$}

\def\pr{{\it Phys. Rev.}\ }
\def\prl{{\it Phys. Rev. Lett.}\ }
\def\pl{{\it Phys. Lett.}\ }
\def\np{{\it Nucl. Phys.}\ }

\def\ijmp{{\it Int. Journ. Mod. Phys.}\ }

\def\cqg{{\it Class. Quantum Grav.}\ }

\def\mnras{{\it Mon. Not. R. Ast. Soc.}\ }

\def\ie{{\it i.e. }}
\def\eg{{\it e.g. }}

\def\f{F(\phi)}

\def\p{\phi}

\def\v{V(\phi)}

\def\l{\cal L}
\begin{document}
\def\bib#1{[{\ref{#1}}]}
\begin{titlepage}
	 \title{ Asymptotic Freedom Cosmology}

 \author{{S. Capozziello$^{1,3}$, R. de Ritis$^{2,3}$, A.A. Marino$^{3,4}$}
\\{\em{\small $^{1}$Dipartimento di Scienze Fisiche, ``E.R. Caianiello'',}}
\\{\em{\small Universit\`a di Salerno, I-84081 Baronissi, Salerno,}}
\\ {\em{\small $^{2}$Dipartimento di Scienze Fisiche, Universit\`{a} di 
Napoli,}}
\\{\em{\small Mostra d'Oltremare pad. 20 I-80125 Napoli,}}
\\ {\em{\small $^{3}$Istituto Nazionale di Fisica Nucleare, Sezione di 
Napoli,}}
\\{\em{\small $^{4}$Osservatorio Astronomico di Capodimonte,}}
\\{\em{\small Via Moiariello 16, I-80131 Napoli, Italy.}}}
	      \date{}
	      \maketitle
	      \begin{abstract}
For a general class of scalar--tensor gravity theories, we discuss how to
recover asymptotic freedom regimes when cosmic time $t\rightarrow\pm\infty$.
Such a feature means that the effective gravitational  coupling
$G_{eff}\rightarrow 0$, while cosmological solutions can asymptotically
assume de Sitter or  power--law behaviours.
In our opinion, through this mechanism, it is possible to cure
some shortcomings in inflationary and in string--dilaton cosmology.

	      \end{abstract}

\vspace{20. mm}
e-mail:\\
 capozziello@vaxsa.csied.unisa.it\\
deritis@axpna1.na.infn.it\\
marino@cerere.na.astro.it
	      \vfill
	      \end{titlepage}
\section{\normalsize \bf Introduction}
In the last decades, people are taking into serious consideration alternative
theories of gravity whose effective actions are more general than
that of Hilbert--Einstein. This approach is mainly due to unification schemes
which must include gravity coherently with the other interactions at a 
quantum level \bib{super}.
From a cosmological point of view, in order to construct more realistic 
models with respect to the standard one, alternative theories are needed
to avoid general relativity shortcomings \bib{brans}, to get inflation
without fine--tuning \bib{la}, to construct perturbation spectra in agreement
with the observed large--scale structure \bib{hwang}.

Anyway, these extended gravity theories (which generally includes 
higher--order terms in curvature invariants or nonminimal couplings
between geometry and one or more  scalar fields \bib{wands})
introduce new features that general relativity, in the Einstein formulation,
does not possess.
Among them, asymptotic freedom seems to emerge as a sort of new
characteristic connected to singularity--free cosmological models or 
inflationary behaviours. Its emergence could be the result of the fact that
gravity is an ``induced'' interaction coming from an average effect of the
other fundamental forces \bib{ottewill}.

However, gravitational asymptotic freedom is just analogue, for example,
with respect to that introduced in strong interactions, 
due to the lack of a full quantum gravity theory.

From a classical point of view, using cosmological models
derived from alternative gravity theories, it is possible to recover
the fact that effective gravitational coupling goes to zero in
asymptotic regimes. 

Taking into account non--Abelian gauge theories (\eg QCD), we have that 
the effective strength of interactions goes to zero at short distances
when the energy diverges \bib{politzer},\bib{ramond}.
The effect was tested by deep--inelastic scattering experiments
\bib{kendall},\bib{bjorken}.

In cosmology, the corresponding feature of ``short distances'' could be
``early times'' (\ie $t\rightarrow -\infty$), where energy diverges
and gravitational coupling has to vary with respect to the present value
in order to recover something similar to the ``running coupling constant''
of QCD \bib{ramond}.
A sort of asymptotic freedom could be recovered also toward the 
future ($t\rightarrow +\infty$) as soon as $G_{eff}\rightarrow 0$.

Several papers have recently been devoted to such a problem.
Among them, some are searching for the effect of gravitational asymptotic
freedom in the large scale structure. In fact, it is possible to show that
a scale--dependent gravitational coupling affects the two--point correlation
function and the Jeans length for structure formation \bib{bertolami}.

It is possible to recover asymptotic freedom also in higher--order gravity
\bib{brandenberger}. In this approach, a limiting curvature appears
at early times and a dynamically consistent picture  is possible at
the Planck epoch \bib{markov}.

Also in string--dilaton gravity, it is possible to recover an 
asymptotic--freedom regime depending on the number of dimensions of the 
theory \bib{elizalde}.

In a previous paper \bib{asfree}, two of the authors faced the problem to
obtain cosmological asymptotic freedom in scalar--tensor theories of gravity.
It was shown that in a Friedman--Robertson--Walker (FRW) flat
spacetime, we can obtain singularity--free solutions where the effective 
gravitational coupling $G_{eff}\rightarrow 0$ for $t\rightarrow -\infty$.
For some of the solutions, we got $G_{eff}\rightarrow G_{N}$ 
for $t\rightarrow +\infty$, where $G_{N}$ is the 
Newton constant. Such features are recovered
if cosmological models satisfy a set of general conditions fixed 
{\it a priori}. In this paper, we generalize those results asking for
asymptotic freedom toward $t\rightarrow\pm\infty$ and searching for those
classes of models, with asymptotic behaviours for the scale factor
$a(t)$ evolving as ${\displaystyle a(t)\simeq e^{\alpha t}}$ and
${\displaystyle a(t)\simeq t^{p}}$.

In Sec.2, we derive the cosmological equations of motion for a general class
of scalar--tensor theories. Sec.3 is devoted to recover a differential 
equation for the gravitational coupling by which it is possible to discuss 
the asymptotic freedom. In Secs.4 and 5, asymptotic freedom is recovered
for exponential and power--law cosmological behaviours. Discussion and
conclusions are drawn in Sec.6. 
  
\section{\normalsize\bf Scalar--tensor FRW cosmology}
We start using the following action
\beq
\label{1}
{\cal A}=\int d^{4}x\sqrt{-g}\left[\f R +\frac{1}{2}
g\umunu \phi\ddemu \phi\ddenu-V(\phi)+{\l}_{m}\right]\,,
\eeq
where $\f$, $\v$ are generic functions of $\p$ and ${\l}_{m}$ is the ordinary
matter Lagrangian density. In a FRW spacetime, the equations of motions are
\beq
\label{2}
H^2+\left(\frac{\dot{F}}{F}\right)H+\frac{k}{a^2}+\frac{\rho_{\p}}{6F}
+\frac{\rho_{m}}{6F}= 0\,,
\eeq
\beq
\label{3}
2\dot{H}+3H^2+2\left(\frac{\dot{F}}{F}\right)H+\frac{\ddot{F}}{F}
+\frac{k}{a^2}-\frac{p_{\p}}{2F}-\frac{p_{m}}{2F}=0\,,
\eeq
\beq
\label{4}
\ddot{\p}+3H\dot{\p}+12F'H^2+6F'\dot{H}+6F'\frac{k}{a^2}+V'=0\,,
\eeq
while the fluid--matter dynamics is given by the conservation equation
\beq
\label{5}
\dot{\rho}_{m}+3H\left(p_{m}+\rho_{m}\right)=0\,;
\eeq
and, as  usual, the state equation is given by
\beq
\label{6}
p_{m}=(\gamma-1)\rho_{m}\,,\;\;\;\;\;\;\;1\leq\gamma\leq 2\,.
\eeq
We have that ${\displaystyle H=\frac{\dot{a}}{a}}$ is the
Hubble parameter, ${\displaystyle \dot{F}=\frac{dF}{d\p}\dot{\p}=F'\dot{\p}}$
is the time derivative of coupling $\f$, 
\beq
\label{7}
\rho_{\p}=\frac{1}{2}\dot{\p}^2+\v\,,\;\;\;\;\;\;
p_{\p}=\frac{1}{2}\dot{\p}^2-\v\,,
\eeq
are, respectively, the energy density and pressure of the scalar field $\p$
while $\rho_{m}$ and $p_{m}$ are the corresponding quantities for ordinary 
matter;
$\gamma$ is a constant.

Usually, the system of differential equation (\ref{2})--(\ref{6}) is solved
for $a(t)$ and $\p(t)$ \bib{noi},
given $\f$, $\v$, $\gamma$, and $k$;
one gets then $a(t)$, $\p(t)$ and $\rho_{\p}$, $p_{\p}$, $\rho_{m}$, $p_{m}$.

On the other side, we can assign the behaviours 
of $a(t)$ and $V(\p(t))$ and look for some particular
behaviour of $F(\p(t))$ and $\p(t)$. However, the
constants $\gamma$ and $k$ have to be given.

We shall face this second approach since we want to ask for asymptotic freedom,
\ie the behaviour of $\f\rightarrow\infty$, for cosmological 
solutions of the forms
${\displaystyle a(t)\sim e^{\alpha t}}$ and ${\displaystyle a(t)\sim t^{p}}$. 
In fact, if asymptotic freedom is recovered, the gravitational coupling
\beq
\label{9}
G_{eff}=-\frac{1}{2\f}\,,
\eeq
goes to zero. Eq.(\ref{9}) is written in Planck's units. The standard 
Newtonian coupling $8\pi G_{N}=1$ is recovered for ${\displaystyle \f=-\frac{1}{2}}$.

\section{\normalsize\bf Recovering gravitational asymptotic freedom}

As first issue, we have to obtain a differential equation for $F(\p(t))$
which analysis could give rise to the  classes 
of models showing asymptotic freedom.
Let us consider the Einstein equations (\ref{2}) and (\ref{3}). Combining 
them, with a little algebra, we get
\beq
\label{10}
\ddot{F}+5H\dot{F}+2\left(3H^{2}+\dot{H}\right)F+V+\frac{4k}{a^2}F-
\frac{(\gamma-2)}{2}\rho_{m}=0\,.
\eeq
Using Eqs.(\ref{5}) and (\ref{6}), we have
\beq
\label{11}
\rho_{m}=\frac{D}{a^{3\gamma}}\,,
\eeq
where $D$ is an integration constant. Then Eq.(\ref{10}) becomes
\beq
\label{12}
\ddot{F}+5H\dot{F}+2\left(3H^{2}+\dot{H}\right)F+V+\frac{4k}{a^2}F-
\frac{(\gamma-2)}{2}\frac{D}{a^{3\gamma}}=0\,.
\eeq
At this point, it is necessary to assign the form of the potential $V$. 
Following the analysis we have already done in the context of
the so called {\it Noether cosmologies} \bib{noi},\bib{cimento},
a simple
but useful choice could be 
\beq
\label{13}
\v=V_{0}\f^{s}\,,\;\;\;\;\v\geq 0\,,
\eeq
$s$ being a free parameter. We consider the case $\f<0$ which has physical 
meaning.
The equation for the coupling $F$ becomes then
\beq
\label{14}
\ddot{F}+5H\dot{F}+2\left(3H^{2}+\dot{H}\right)F+V_{0}F^{s}+\frac{4k}{a^2}F-
\frac{(\gamma-2)}{2}\frac{D}{a^{3\gamma}}=0\,.
\eeq
It is worthwhile to note that such an equation can be invariant under time 
reversal ($t\rightarrow -t$) for suitable choices of the parameters $V_{0}$
and $s$ in the sense  
that the functions  $F(t)$,
$a(t)$, and $F(-t)$, $a(-t)$  satisfy the same differential equation.
This feature is interesting to get asymptotic freedom toward 
$t\rightarrow\pm\infty$. 

\section{\normalsize\bf The exponential behaviour}

Let us consider, for the scale factor of the universe $a(t)$, an asymptotic
behaviour of the form
\beq
\label{15}
a(t)=a_{0}e^{\alpha t}\,,\;\;\;\;
\alpha>0\,,\;\;\;\;
t\gg 0\,,
\eeq
where $\alpha$ is a constant. Considerations similar to that below
will hold for $t\ll 0$.

Being $H=\alpha$, $\dot{H}=0$, we get
\beq
\label{16}
\ddot{F}+5\alpha\dot{F}+6\alpha^{2}F+V_{0}F^{s}+
\frac{4k}{a_{0}^2}e^{-2\alpha t}F-
\frac{(\gamma-2)}{2}\frac{D}{a_{0}^{3\gamma}}e^{-3\gamma\alpha t}=0\,.
\eeq
A simple choice is assuming a FRW spacetime with $k=0$. Furthermore,
we are interested in asymptotic regime so that 
ordinary matter term has no influence and then it can be discarded.
The asymptotic equation for the coupling becomes
\beq
\label{17}
\ddot{F}+5\alpha\dot{F}+6\alpha^{2}F+V_{0}F^{s}=0\,,
\eeq
which depends on the free parameter $s$. This ordinary differential equation is
quite general and, in principle, holds any time we get an asymptotic
de Sitter behaviour in a nonminimally coupled theory.

\subsection{\normalsize\bf The case $s=0$}

If $s=0$ and $V_{0}\geq 0$, the general solution of Eq.(\ref{17}) is
\beq
\label{18}
F(t)=A_{1}e^{-2\alpha t}+A_{2}e^{-3\alpha t}-\frac{V_{0}}{6\alpha^{2}}\,,
\eeq
where $A_{1,2}$ are integration constants.
It is interesting to see that, by using the definition (\ref{9}),
\beq
\label{19}
F\longrightarrow -\frac{V_{0}}{6\alpha^2}\,,\;\;\;
G_{eff}\longrightarrow \frac{3\alpha^2}{V_{0}}\,,
\eeq
and
\beq
\label{20}
\frac{\dot{F}}{HF}\longrightarrow 0\,,\;\;\;
\frac{\dot{G}_{eff}}{HG_{eff}}\longrightarrow 0\,,
\eeq
for $t\rightarrow +\infty$, so that it is possible to recover standard
gravity in such a regime.
With respect to the considerations in \bib{asfree}, the situation here
presented
is new and, by it, the time--coupling variation $\dot{F}/F$ is
compared with cosmological evolution $H$. 

Substituting solution (\ref{18}) into  the Klein--Gordon Eq.(\ref{4}),
(with $k=0$), we get the behaviour of the field $\p(t)$
\beq
\label{21}
\p(t)=A_{3}e^{-3\alpha t}+\p_{0}\,,
\eeq
so that the coupling is
\beq
\label{22}
\f=A_{1}\left(\frac{\p-\p_{0}}{A_{3}}\right)^{2/3}
+A_{2}\left(\frac{\p-\p_{0}}{A_{3}}\right)
-\frac{V_{0}}{6\alpha^2}\,.
\eeq
Also here, $A_{3}$ is an integration constant. Standard gravity is recovered 
since, for $t\gg 0$,
\beq
F\rightarrow -\frac{V_{0}}{6\alpha^2}<0\,.
\eeq

\subsection{\normalsize\bf The case $s=1$}
Another interesting choice is
\beq
\label{23}
s=1\,,\;\;\;\;\;F<0\,,\;\;\;\;V_{0}<0\,.
\eeq
The solution of Eq.(\ref{17}) is
\beq
\label{24}
F(t)=\exp\left(-\frac{5\alpha}{2}t\right)
\left[A_{1}\exp\left(\alpha\sqrt{\frac{1}{4}-\frac{V_{0}}{\alpha^2}}t\right)
+A_{2}\exp\left(-\alpha\sqrt{\frac{1}{4}-\frac{V_{0}}{\alpha^2}}t\right)
\right]\;.
\eeq
Asymptotically, for $t\rightarrow +\infty$, we have
\beq
\label{25}
F\longrightarrow -\infty\,,\;\;\;\;\;\;
G_{eff}\longrightarrow 0\,,
\eeq
if
\beq
\label{26}
\frac{|V_{0}|}{6\alpha^2}>1\,,\;\;\;\;\;\;A_{1}<0\,,
\eeq
and, always for $t\rightarrow +\infty$,
\beq
\label{27}
F\longrightarrow 0\,,\;\;\;\;\;\;
G_{eff}\longrightarrow \infty\,,
\eeq
if
\beq
\label{28}
\frac{|V_{0}|}{6\alpha^2}<1\,,\;\;\;\;\;\;A_{2}<0\,.
\eeq
In the case (\ref{25}), we recover the asymptotic  freedom. 
We have that
\beq
\label{29}
\frac{\dot{F}}{HF}\longrightarrow -\left(\frac{5}{2}-
\sqrt{\frac{1}{4}-\frac{V_{0}}{\alpha^2}}\right)\;;
\;\;\;\;\;\;
\frac{\dot{G}_{eff}}{HG_{eff}}\longrightarrow \left(\frac{5}{2}-
\sqrt{\frac{1}{4}-\frac{V_{0}}{\alpha^2}}\right)\;.
\eeq
If conditions (\ref{28}) holds, the situation is analogue to the case $s=0$.
If
\beq
\label{30}
1<\frac{|V_{0}|}{6\alpha^2}<2\,,\;\;\;\;\;\;
A_{1}<0\,,
\eeq
we get
\beq
\label{31}
\f=\xi_{0}\left(\p-\p_{0}\right)^{2}\,,
\eeq
where
\beq
\label{32}
\xi_{0}=-\frac{\left(\frac{7}{2}+\sqrt{\frac{1}{4}-\frac{V_{0}}{\alpha^2}}
\right)\left(-\frac{5}{2}+\sqrt{\frac{1}{4}-\frac{V_{0}}{\alpha^2}}
\right)}{48\left(2+\frac{V_{0}}{6\alpha^2}\right)}\,.
\eeq
Of course, in this case, the potential assume a form similar to the coupling
(\ref{31}).
Similar analysis can be performed for other values of $s$. 
The main point is that, by imposing an asymptotic de Sitter behaviour
for the scale factor $a(t)$, we can recover asymptotic freedom in the case of
attractive  gravity toward $t\rightarrow  +\infty$.
It straightforward to see that analogous results hold for $t\rightarrow
-\infty$ and $a\simeq e^{-\alpha t}$, $\alpha >0$, that is one can have 
asymptotic freedom for infinitely negative $t$.

\section{\normalsize\bf The power--law behaviour}
Let us now assume an asymptotic power--law behaviour for the scale--factor of
the universe, that is
\beq
\label{33}
a(t)\sim t^{p}\,,\;\;\;\;p>0\,,\;\;\;\;t\gg 0\,.
\eeq
Eq.(\ref{14}) becomes
\beq
\label{34}
\ddot{F}+\frac{5p}{t}\dot{F}+\frac{1}{t^2}\left[2p(3p-1)+
\left(\frac{4kt_{0}^{2}}{a_{0}^2}\right)
\left(\frac{t_{0}}{t}\right)^{2p-2}\right]F+V_{0}F^{s}-\frac{(\gamma-2)}{2}
\frac{D}{a_{0}^{3\gamma}}\left(\frac{t_{0}}{t}\right)^{3\gamma p}=0\,,
\eeq
being
\beq
\label{35}
a(t)=a_{0}\left(\frac{t_{0}}{t}\right)^{p}\,,\;\;\;\;\;
H=\frac{p}{t}\,,\;\;\;\;
\dot{H}=-\frac{p}{t^{2}}\,.
\eeq
Eq.(\ref{34}) can be a Fuchs equation \bib{ince} depending on the 
values of $k$ and $p$. As before, a simple choice is  $k=0$.

\subsection{\normalsize\bf The case $s=0$}
If $s=0$, for $t\rightarrow +\infty$, it is easy to see that
\beq
\label{36}
\frac{(\gamma -2)}{2}\frac{D}{a_{0}^{3\gamma}}
\left(\frac{t_{0}}{t}\right)^{3\gamma p}\ll V_{0}\,,
\eeq
and then Eq.(\ref{34}) reduces to
\beq
\label{37}
\ddot{F}+\frac{5p}{t}\dot{F}+\frac{2p(3p-1)}{t^2}F+V_{0}=0\,.
\eeq
The homogeneous associated equation (\ie $V_{0}=0$) is a totally 
Fuchsian equation whose general solution is of the form
\beq
\label{38}
F(t)\sim t^{r}+B_{1}t^2+B_{2}t+B_{3}\,,
\eeq
where $r$ has the values $r_{1}=1-3p$ or $r_{2}=-2p$, so that
\beq
\label{40}
F(t)=A_{1}t^{1-3p}+A_{2}t^{-2p}+B_{1}t^{2}+B_{2}t+B_{3}\,.
\eeq
Substituting into Eq.(\ref{37}), we get
\beq
\label{41}
B_{1}=-\frac{V_{0}}{2p(3p-1)+2}\,,\;\;\;\;\;
B_{2}=B_{3}=0\,.
\eeq
The parameter $p$ can assume any non--negative real value. The final solution
is
\beq
\label{42}
F(t)=A_{1}t^{1-3p}+A_{2}t^{-2p}-\frac{V_{0}t^2}{2p(3p-1)+2}\,.
\eeq
Asymptotically, for $t\rightarrow +\infty$, we have
\beq
\label{43}
F\longrightarrow -\infty\,;\;\;\;\;\;G_{eff}\longrightarrow 0\,,
\eeq
which means the recovering of asymptotic freedom.

\subsection{\normalsize\bf The case $s=1$}
Another choice can be $s=1$ for $F<0$ which implies $V_{0}<0$.
Eq.(\ref{14}) is
\beq
\label{44}
\ddot{F}+\frac{5p}{t}\dot{F}+\left[\frac{2p(3p-1)}{t^2}+V_{0}
\right]F
-\frac{(\gamma-2)}{2}
\frac{D}{a_{0}^{3\gamma}}\left(\frac{t_{0}}{t}\right)^{3\gamma p}=0\,,
\eeq
whose homogeneous associated equation is
\beq
\label{45}
\ddot{F}+\frac{5p}{t}\dot{F}+\left[\frac{2p(3p-1)}{t^2}+V_{0}\right]F=0\,,
\eeq
which is not a Fuchs equation for $t\rightarrow +\infty$.
We can search for a solution of the form
\beq
\label{46}
F(t)\sim e^{\beta t}t^{m}\Sigma_{n=0}^{\infty}\frac{f_{n}}{t^{n}}\,,
\eeq
where $\beta$, $m$, $n$ and $f_{n}$ are arbitrary constants \bib{ince}.
Inserting (\ref{46}) into (\ref{45}), we find
\beq
\label{47}
F(t)=t^{-(5p)/2}\left[A_{1}e^{\sqrt{|V{_{0}}|}}
+A_{2}e^{-\sqrt{|V{_{0}}|}}\right]\,,
\eeq
and then, for $t\rightarrow +\infty$,
\beq
\label{48}
F\longrightarrow -\infty\,,\;\;\;\;\;\;
G_{eff}\longrightarrow 0\,,
\eeq
if $A_{1}<0$ and $p$ is real and non--negative.

The behaviour of $\p(t)$ and the dependence of $F$ by $\p(t)$ 
can be recovered 
as above by taking into account the Klein--Gordon Eq.(\ref{4}). 
Similar considerations hold for the other values of $s$.

\section{\normalsize\bf Discussion and conclusions}
Asymptotic freedom seems to be a common feature for general classes
of scalar--tensor theories of gravity, \ie when 
the gravitational coupling is assumed to depend on a scalar field.

Furthermore,  we have shown that it can be recovered for 
$t\rightarrow\pm\infty$ so that, as supposed for example, in string--dilaton
cosmology \bib{string}, it seems that a sort of asymptotic symmetry 
exists in several classes of cosmological models.
It strictly depends on the relative signs of the parameters inside
the scalar--field potentials and couplings, but many kinds of cosmological
behaviours (\ie ${\displaystyle a(t)\sim e^{\alpha t}}$ or
${\displaystyle a(t)\sim t^{p}}$) of physical interest can exhibit such a
feature.

The main point is to obtain from the dynamical equations of a 
certain class of cosmological models, an ordinary differential equation
which gives the behaviour of the gravitational coupling $F(\p(t))$.

Furthermore, some models allow to recover the standard Einstein gravity
(at least an attractive behaviour for the interaction) as soon as
$G_{eff}\rightarrow constant$. In this sense, general relativity is not
ruled out by the presence of asymptotic freedom but it is a complementary
feature for particular classes of models.

As discussed also in \bib{asfree}, this situation can be read as a classical
analogue of QCD toward a full theory in which gravity is an interaction
induced by the other non--gravitational fundamental forces. 

\vspace{5. mm}

\begin{centerline}
{\bf REFERENCES}
\end{centerline}
\begin{enumerate}

\item\label{super}
A. Zee, \prl {\bf 42} (1979) 417.\\
L. Smolin, \np {\bf B 160} (1979) 253.\\
N.D. Birrell and P.C.W. Davies, {\it Quantum Fields in Curved Space}
(Cambridge Univ. Press, Cambridge, 1986).\\
G. Vilkovisky, \cqg {bf 9} (1992) 895.
\item\label{brans}
D.W. Sciama, \mnras {\bf 113} (1953) 34.\\
C. Brans and R.H. Dicke, \pr {\bf 124} (1961) 965.
\item\label{la}
C. Mathiazhagen and V.B. Johori, \cqg {\bf 1} (1984) L29.\\
D. La  and P.J. Steinhardt,  {\it Phys. Rev. Lett.} {\bf 62} (1989) 376. \\
D. La,  P.J. Steinhardt  and E.W. Bertschinger, 
{\it Phys. Lett.} {\bf B 231} (1989) 231. \\
P.J. Steinhardt and F.S. Accetta, \prl {\bf 64} (1990) 2470.\\
A.D. Linde  \pl {\bf B 238} (1990) 160.\\
R. Holman, E.W. Kolb, S. Vadas  and Y. Wang,  \pr {\bf D 43} (1991) 995.
\item\label{hwang}
J. Hwang, \cqg {\bf 7} (1990) 1613.\\
J. Hwang, \pr {\bf D 42} (1990) 2601.
\item\label{wands}
D. Wands, \cqg {\bf 11} (1994) 269.
\item\label{ottewill}
J. Barrow and A. C. Ottewill, {\it J. Phys. A: Math. Gen.} {\bf 16} (1983)
2757.
\item\label{politzer}
H.D. Politzer, {\it Phys. Rep.} {\bf 14} (1974) 129.
\item\label{ramond}
P. Ramond, {\it Field Theory: A Modern Primer} Addison--Wesley Pub. Co.
Menlo Park (Ca) (1989).
\item\label{kendall}
H. Kendall, {\it Proc. Vth Int. Symp. Electron and Photon Interactions
at High Energies}, Cornell Univ. (1971).
\item\label{bjorken}
J.D. Bjorken, \pr {\bf 179} (1969) 1547.
\item\label{bertolami}
O. Bertolami, J.M. Mour\~ao and J. P\'erez--Mercader,
\pl {\bf B 311} (1993) 27.\\
O. Bertolami \pl {\bf B 186} (1987) 161.
\item\label{brandenberger}
R. Brandenberger, {\it Proc. of the 6'th Quantum Gravity Seminar, Moskow}
ed. V. Berezin et al. World Scientific, Singapore, 1996.
\item\label{markov}
M.A. Markov, in {\it The Very Early Universe}, Proceedings of the
Nuffield Workshop, eds.G.W. Gibbons, S.W. Hawking, S.T.C. Siklos,
Cambridge Univ. Press., Cambridge (1982).\\
M.A. Markov, {\it Phys. Uspekhi} {\bf 37} (1994) 57.
\item\label{elizalde}
E. Elizalde and S.D. Odintsov, \pl {\bf B 347} (1995) 211.
\item\label{asfree}
S. Capozziello and R. de Ritis, \pl {\bf A 208} (1995) 181.
\item \label{noi}
S. Capozziello and R. de Ritis, \pl {\bf A 177} (1993) 1.\\
S. Capozziello and R. de Ritis, \cqg {\bf 11} (1994) 107.
\item\label{cimento}
S. Capozziello, R. de Ritis, C. Rubano, and P. Scudellaro,
{\it La Rivista del N. Cimento} {\bf 4} (1996) 1.
\item\label{ince}
E.L. Ince, {\it Ordinary Differential Equations} 
Dover Publ. Inc., New York (1956).
\item\label{string}
M.B. Green, J.H. Schwarz, E. Witten {\it Superstring Theory} 
Cambridge Univ. Press, Cambridge (1987).\\
A.A. Tseytlin and C. Vafa, \np {\bf B 372} (1992) 443.\\
M. Gasperini, J. Maharana and G. Veneziano, \pl {\bf B 272} (1991) 277.\\
S. Capozziello and R. de Ritis \ijmp {\bf 2D} (1993) 373.
\end{enumerate}
\vfill

\end{document}